\documentclass[aps,prx,twocolumn,amsmath,nofootinbib,longbibliography,amssymb,superscriptaddress,10pt]{revtex4-2}
\usepackage[margin=1in]{geometry}
\usepackage[english]{babel}
\usepackage{etoolbox}
\usepackage[shortlabels]{enumitem}
\usepackage[scr=boondox]{mathalpha}   % slightly sloped
           
\usepackage{amsmath,amsthm,amssymb,amsfonts}
\usepackage{stmaryrd}
\usepackage{accents}
\usepackage{bm}
\usepackage{leftindex}
\usepackage{dsfont} % double stroked numerals
\usepackage{transparent}
\usepackage[linkcolor=blue,colorlinks=true]{hyperref}
\usepackage{graphicx}
\usepackage[caption=false]{subfig}
\graphicspath{
    {figures/}
}

\docsvlist{A,B,C,D,E,F,G,H,I,J,K,L,M,N,O,P,Q,R,S,T,U,V,W,X,Y,Z}

\docsvlist{A,B,C,D,E,F,G,H,I,J,K,L,M,N,O,P,Q,R,S,T,U,V,W,X,Y,Z}

\docsvlist{j}

\docsvlist{R,k,p,q,r,g,x,y}

\def \ve {h} % placeholder for random field notation ###############
\def \bra {\langle}
\def \ket {\rangle}

\begin{document}

\title{Infinitely Stable Disordered Systems on Emergent Fractal Structures}

\author{Andrew C. Yuan}
\affiliation{Department of Physics, Stanford University, Stanford, CA 94305, USA}
\affiliation{Condensed Matter Theory Center and Joint Quantum Institute, Department of Physics, University of Maryland, College Park, Maryland 20742, USA}

\author{Nick Crawford}
\affiliation{Department of Mathematics, The Technion, Haifa, Israel}

\date{\today}
\begin{abstract}
    In quenched disordered systems, the existence of ordering is generally believed to be only possible in the weak disorder regime (disregarding models of spin-glass type). 
    In particular, sufficiently large random field is expected to prohibit any finite temperature ordering. 
    % for models in which the order parameter couples to a random field, ordering at finite temperatures is prohibited, provided the disorder strength is sufficiently large.
    % In particular, for models in which the order parameter couples to a random field, ordering at finite temperatures is prohibited, provided the disorder strength is sufficiently large.
    Here, we show that this is not necessarily true. We provide physically motivated  examples of systems in which disorder induces an ordering that is \textit{infinitely stable} in the sense that: (1) there exists ordering at arbitrarily large disorder strength and (2) the transition temperature remains, asymptotically, nonzero in the limit of infinite disorder.
    The ordering is spatially localized on the boundary of a disorder-induced, emergent percolating fractal structure.
    %Interestingly, the ordering lives only on the boundary of a (disorder-induced) emergent percolating fractal structure.
    % The disorder-induced-ordering depends the existence of a percolating  
    %on the boundary of an emergent 
    % fractal structure in $\mathbb{Z}^d$, whose boundary provides a volume-extensive contribution in the thermodynamic limit.
    The examples we give are most naturally described when the spatial dimension $d \ge 3$, but can also be formulated when $d=2$, provided that the underlying graph is non-planar. %(e.g., $\mathbb{Z}^2$ sites with nearest and next-nearest neighbor interactions).
\end{abstract}
\maketitle

\begin{figure}[ht]
\centering
\includegraphics[width=.8\columnwidth]{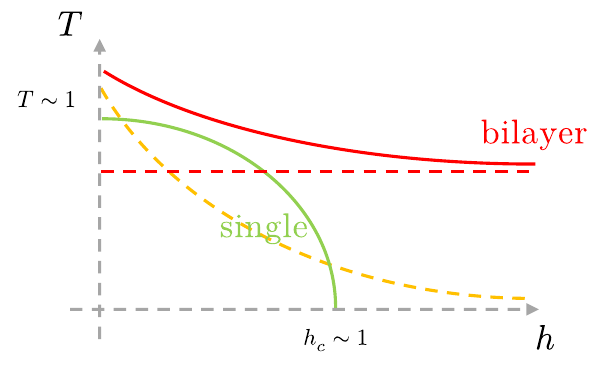}
\caption{Schematic Phase Diagram. The green line sketches the transition temperature $T_c(\ve)$ of \textit{single layer} models with disorder strength $\ve$, e.g., the random-field Ising  and XY models \cite{aharony1978tricritical,aharony1978spin,bray1985scaling}, the RFO(2) model \cite{dotsenko19812d,dotsenko1982spin,crawford2024random,crawford2013random,crawford2014random}.
If $\ve \to 0$, then $T_c(\ve)\sim 1$ is on the scale of the XY coupling strength $\kappa=1$ (in 2D, there is debate on whether LRO or quasi-LRO occurs below $T_c$ for the RFO(2) model \cite{crawford2024random}).
For these models, there exists a critical threshold $\ve_c \sim 1$, beyond which there is no finite temperature ordering.
In comparison, this Letter introduces a \textit{bilayer} model with $T_c(\ve)$ sketched by the red line.
If $\ve \to \infty$, then $T_c(\ve)$ is asymptotically nonzero (dashed red line).
For completeness, the dashed orange line denotes an intermediate scenario where there is finite temperature ordering at large disorder, but $T_c(\ve)$ is asymptotically zero.
}
\label{fig:schematic-phase}
\end{figure}

It's generally recognized that ordering can exist in the presence of quenched disorder.
However, the existence of randomness is usually an undesirable effect which disrupts coherence and thus lowers the transition temperature.
The Imry-Ma phenomenon \cite{imry1975random} provides an extreme example; in the presence of random fields, the phase transition of spin systems can be destroyed in low dimensions independent of the field strength  ($d=2, 4$ for discrete, continuous symmetries, respectively).
The statement has, in fact, been rigorously demonstrated for the random field Ising and XY models \cite{aizenman1989rounding,aizenman1990rounding}, along with their quantum versions \cite{greenblatt2009rounding,aizenman2012proof}. 
More recent work even provides quantitative bounds on the decay of correlations in ground states (GS) \cite{ding2021exponential,aizenman2020exponential}.
In these systems, even in higher dimensions so that Imry-Ma does not apply, there generally is a critical threshold $\ve_c$ in disorder strength above which any finite temperature ordering is destroyed \cite{aharony1978tricritical,aharony1978spin,bray1985scaling} (see Fig. \ref{fig:schematic-phase}).

The existence of a critical threshold $\ve_c$ 
% above which there is no ordering at nonzero temperatures 
is also true in lesser known examples where the disorder \textit{generates} ordering \cite{dotsenko19812d,dotsenko1982spin,crawford2024random,crawford2013random,crawford2014random}. 
In these models, the random field disorder acts along a submanifold of the spin space and,  while Imry-Ma can prohibit ordering along the submanifold, ordering occurs in the direction perpendicular to the submanifold.  
% Upon inspection, small amplitude spatial fluctuations of the random field seem to be crucial for this effect. 
For example, in the RFO(2) model \cite{crawford2013random,crawford2014random,crawford2024random},  the random field disorder acts only in the X direction of the XY spins and the spins can still order in the Y direction at finite temperatures, provided that the disorder is sufficiently weak.
At larger disorder strength, this ordering is destroyed as the energetic cost for not aligning with the random field becomes prohibitive \cite{aharony1978spin}.
From this discussion, with the notable exception of spin glass ordering, one might speculate that spontaneous symmetry breaking in disordered systems is only possible within the weak disorder regime.
% For concreteness, we coin these \textit{single layer} models since they disorder usually acts onsite in addition to a standard clean model (Ising or XY). 

Contrary to the last hypothesis, in this Letter we provide a family of models which exhibit \textit{infinitely stability} (our terminology) with respect to disorder strength $\ve$, i.e., exhibit long-range ordering (LRO) below a finite  $T_\text{LRO}$ which is \textit{independent} of the disorder strength in the limit $\ve \to \infty$ (see Fig. \ref{fig:schematic-phase}).  
Our consideration of these systems is primarily motivated by recent experiments on twisted bilayer superconductors (SC) \cite{zhao2023time,tummuru2022josephson,can2021high,yuan2024absence,yuan2023exactly,yuan2023inhomogeneity}, though it is worth mentioning that, after some reasonable simplifications, our examples resemble models studied in the context of frustrated antiferromagnets \cite{henley1987ordering,chandra1990ising,chubukov1992order,Henley1989}.  
In the latter cases, random dilution sites in the underlying lattice can lead to \textit{anticollinear} ordering, as observed in Cd$_{1-p}$Mn$_p$Te \cite{giebultowicz1986monte,giebultowicz1988neutron,larson1988theory}.
In experiments, the geometry of the underlying lattice is two dimensional, but the setup extends straightforwardly to higher dimensions (which may model multi-component SCs with random inter-component coupling \cite{carlstrom2011length,bojesen2014phase,grinenko2021state}).

The reason behind this stable form of ``order-by-disorder'' is the existence of a pair of emergent, complementary site percolation processes at parameter $p=1/2$.
% which partitions the lattice into a pair of random subgraphs.  
Crucially, the transition temperature in the limit $\ve \to \infty$ is nonzero only if both random environments percolate, i.e., $p_c^\text{site}(G) < 1/2$ where $G$ denotes the underlying graph of the model.
% sites of the lattice together with edges defined by the intra-layer couplings.
If $G=\dZ^d,d\ge 3$, $p_c^\text{site}(G) < 1/2$ by a monotonicity argument\footnote{This follows for $\dZ^3$ by comparing with site percolation on the triangular lattice obtained by projecting in the $(1,1,1)$ direction \cite{CampaninoRusso85}.} and thus there is an infinitely stable phase transition in our examples.
For 2D systems, the situation is more complicated.
Since\footnote{In fact, this holds on any planar graph \cite{grimmett2022hyperbolic}} $p_c^\mathrm{site}(\dZ^2) \ge 1/2$, the previous rule-of-thumb fails.
Nevertheless,  the introduction of next-nearest neighbor (NNN) couplings as well as nearest neighbor (NN) couplings drives  $p_c^\text{site}(\tilde{\dZ}^2) < 1/2$, seen again by monotonicity.
In this case, we show that the system on $\tilde{\dZ}^2$ is at least \textit{infinitely quasi-stable}, i.e., exhibits quasi-LRO below a finite transition temperature $T_\text{QLRO}$ independent of $\ve$.
Whether 2D systems can be infinitely stable is an open question.  
A detailed discussion appears in Sec. (\textit{2D behavior}).  

\medskip
\noindent

\textit{The model} --- In this section, we introduce the models precisely and give an intuitive explanation for the stability phenomenon described above. Further details are provided elsewhere \cite{detailed}.

Given a general graph $G=(V, E)$ consisting of sites $V$ and edges $E,$ consider the following classical Hamiltonian $\sH$, 
\begin{equation}
    \label{eq:H}
    \sH_\alpha(\theta^\pm) = \sum_{\ell=\pm} \sH^\text{XY} (\theta^\ell) -\ve \sum_{\br} \cos (\phi(\br) -\alpha(\br)).
\end{equation}
Here each site $\br$ in $V$ has two \textit{layers} of XY spin degrees of freedom $\theta^\pm (\br)$. 
The first term denotes the Hamiltonian for a pair of identical and independent ferromagnetic XY models of the spin configurations $\theta^\pm$ with \textit{intra-layer} interactions characterized by the edges $e$ of graph $G$.
The second term couples the two XY layers by shifting the phase differences field $\phi(\br) \equiv \theta^+(\br) -\theta^-(\br)$ by a random phase $\alpha(\br) = 0,\pi$ with coupling strength $\ve$.
%by an \textit{inter-layer} interaction dependent only on the phase difference $\phi \equiv \theta^+ -\theta^-$ shifted 
For conceptual clarity, we shall assume that $\alpha(\br)$ for different lattice sites $\br$ are uncorrelated and equally probable to be $0$ or $\pi$ (choosing the distribution of $\alpha(\br)$ uniform over $[0, 2\pi)$ leads to an interesting variation \cite{detailed}).

Physically, when the graph $G$ is 2D (e.g., $\dZ^2$ or $\tilde{\dZ}^2$), $\sH_\alpha$ is motivated by twisted bilayer SCs with interlayer random-in-sign Josephson coupling \cite{zhao2023time,tummuru2022josephson,can2021high,yuan2024absence,yuan2023exactly,yuan2023inhomogeneity}.
In this setting, each SC layer is characterized by an independent XY model with in-plane SC stiffness $\kappa=1$, while the inter-layer Josephson coupling is given by $J(\br) \cos \phi(\br)$ where $J(\br) = \pm \ve$ (depending on whether $\alpha(\br) = 0,\pi$) are inhomogeneities in the system with a disorder-average $\bar{J} = 0$ and standard deviation $\ve$.
Such a system occurs when the two layers are twisted relative to each other at a critical angle so that the lowest order Josephson coupling vanishes on average, i.e. $\bar{J} =0$ (e.g., in experiments this occurs at twist  $= \pi/4$ in BSCCO \cite{zhao2023time}).  
Twist angle disorder or other inhomogeneities can then induce a random-in-sign coupling.

% \begin{figure}[ht]
% \centering
% \includegraphics[width=0.6\columnwidth]{effective-unit-cell.pdf}
% \caption{Effective Unit Cell. The underlying lattice denotes a realistic system with bare in-plane coupling $\kappa$, disorder correlation $\xi$ and bare disorder strength $\ve$. Each effective unit cell is of linear size $\xi$ (dashed blue lines) so neighboring cells are essentially uncorrelated.
% }
% \label{fig:effective-unit-cell}
% \end{figure}

Note that the model in Eq. \eqref{eq:H} assumes uncorrelated disorder, and thus only depends on the in-plane coupling $\kappa$ ($=1$) and disorder strength $\ve$. 
To justify our emphasis on the strong disorder regime, observe that in a realistic setting there exists an additional disorder correlation length $\xi$ so that the system is described by parameters $(\kappa, \ve, \xi)$.
One way of heuristically mapping the more realistic setup to the uncorrelated model in Eq. \eqref{eq:H} is by a simple block coarse graining of the microscopic lattice.  
The \textit{effective} unit cell has linear size $\xi$ so that neighboring cells are essentially uncorrelated in disorder.  
Since the inter-layer interaction acts on-site, the \textit{effective} disorder strength is mapped to $ \ve \xi^d$ while the in-plane coupling strength scales as $ \kappa \xi^{d-1}$ since it is a boundary effect.
In particular, if $\xi  \gg \kappa/\ve $, then the system would be in the strong disorder regime.

\begin{figure}[ht]
\subfloat[\label{fig:obd-a}]{%
  \centering
  \includegraphics[width=0.33\columnwidth]{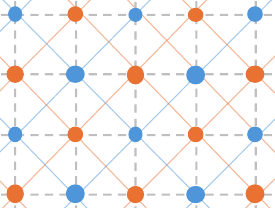}
}
\subfloat[\label{fig:obd-b}]{%
  \centering
  \includegraphics[width=0.32\columnwidth]{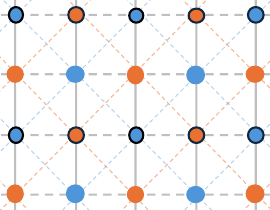}
}
\subfloat[\label{fig:obd-c}]{%
  \includegraphics[width=0.32\columnwidth]{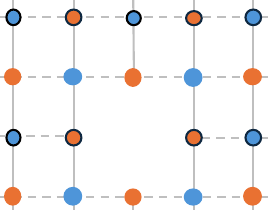}
}
\caption{Order by Disorder \cite{Henley1989}.  
(a) denotes an XY model on $\tilde{\dZ}^2$, with ferromagnetic (dashed lines) NN coupling and antiferromagnetic (solid lines) NNN coupling. 
The lattice is partitioned into even and odd sites (blue and orange). 
(b) After gauge transform on sites circled in black, the NNN couplings becomes ferromagnetic, while horizontal/vertical nn couplings are ferro/antiferromagnetic. (c). Omitting the NNN ferromagnetic couplings, a single site is removed from the lattice, causing an imbalance of ferro/antiferromagnetic couplings at neighboring sites.
}
\label{fig:obd}
\end{figure}

At weak disorder the physics in our examples resemble the frustrated antiferromagnets alluded to above,  \cite{henley1987ordering,chandra1990ising,chubukov1992order,Henley1989}. %(though only in the weak disorder regime). 
In particular,
Ref. \cite{Henley1989} considers an XY model on $\tilde{\dZ}^2$, with ferromagnetic nearest neighbor (NN) coupling $|\kappa_\text{1}|$ and antiferromagnetic NNN coupling $|\kappa_\text{2}|$ as illustrated in Fig. \ref{fig:obd-a}.
By partitioning the lattice into even and odd sublattices, the system describes a bilayer model.
% with antiferromagnetic intra-layer coupling $|\kappa_\text{NNN}|$ and ferromagnetic inter-layer coupling $|\kappa_\text{NN}|$.
Under the gauge transformation ($\theta_i \mapsto \theta_i +\pi$ for sites $i$ in black in Fig. \ref{fig:obd-b}), the system is equivalent to a bilayer XY model with an inter-layer coupling $\pm |\kappa_\text{1}|$ whose sign is orientationally dependent.  In this pure system and for weak inter-sublattice coupling, $|\kappa_\text{1}| \ll |\kappa_\text{2}|$,  the gauge-transformed GSs are presumably ferromagnetically ordered within each sublattice, but the relative orientation between sublattices is yet to be determined.  
Note that in this case the NN interactions (delicately) cancels each others effects at each site.  However, if the sites were randomly diluted (see Fig. \ref{fig:obd-c}),
%For every site $i$, there are an equal number of ferro/antiferromagnetic inter-layer couplings, 
% a small randomly chosen fraction of sites (see Fig. \ref{fig:obd-c}), 
then the NNs of removed sites have an imbalance between the number of ferro- and antiferromagnetic inter-layer couplings.  
This, in turn, leads to an effective random-in-sign inter-layer coupling with weak disorder $|\kappa_\text{1}|$.
Ref. \cite{Henley1989} then argues that the spins between sublattices prefer to align perpendicularly in the weak disorder regime, consistent with Ref. \cite{yuan2023exactly,yuan2023inhomogeneity,crawford2013random,crawford2014random,crawford2024random}.

% ======================
\begin{figure}[ht]
\subfloat[\label{fig:schematic-lattice}]{%
  \centering
  \includegraphics[width=0.54\columnwidth]{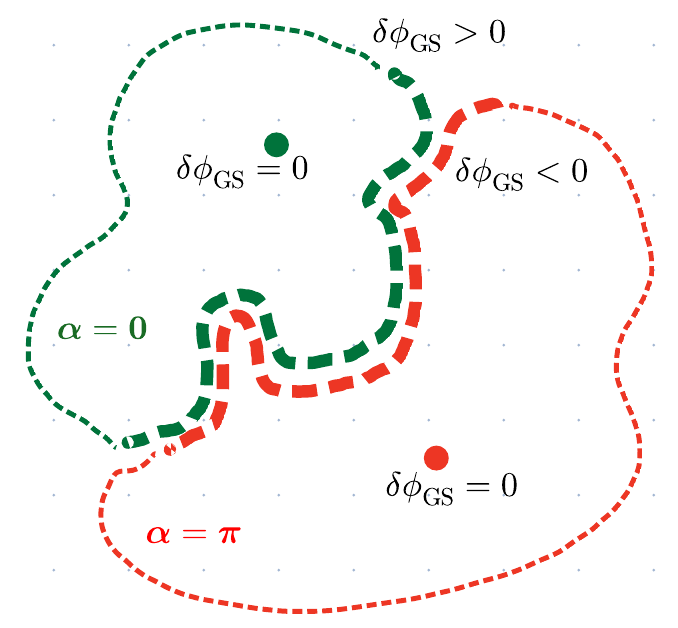}
}
\subfloat[\label{fig:schematic-spin}]{%
  \centering
  \includegraphics[width=0.45\columnwidth]{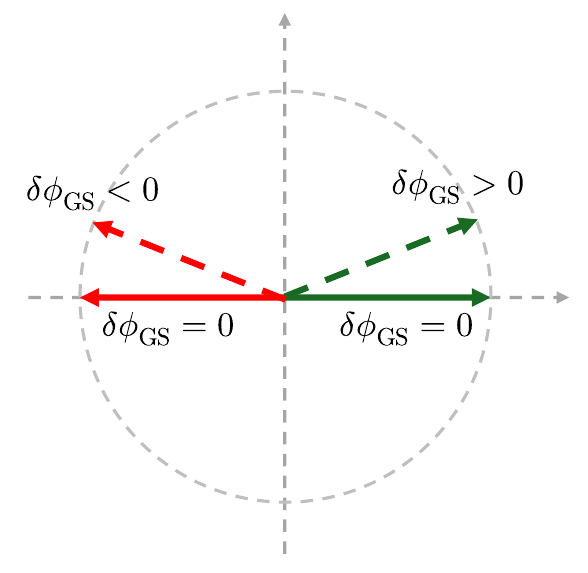}
}
\caption{GS Schematic $\phi\sim^\text{s} +\pi/2$.  In the strong disorder regime, (a) shows a typical disorder realization of $\alpha=0,\pi$ clusters where the green and red dashed lines denote the cluster boundary. By Eq. \eqref{eq:GS}, the slanting $\delta \phi_\text{GS} = \phi -\alpha$ vanishes within the interiors of the clusters (e.g., the colored dots deep within each cluster), but satisfies $\delta \phi_\text{GS}\gtrless 0$ at the boundary. (b) shows the corresponding GS $\phi$ in spin space $\dS^1$. In the interior of clusters (solid lines), $\phi =\alpha$, but at the boundary (dashed lines), there is uniform slanting in the $+\pi/2$ direction.
}
\label{fig:GS}
\end{figure}
\textit{Ground States} --- In the limit of weak disorder, Ref. \cite{yuan2023inhomogeneity} shows that there are $\dZ_2$ degenerate (disorder-averaged) GSs\footnote{In the physical bilayer setup, $\phi \mapsto -\phi$ corresponds to time-reversal symmetry (TRS) and thus the $\dZ_2$ degenerate ground states indicates TRS breaking.}  characterized by a phase difference $\phi \sim^\text{w} \pm \pi/2$ (perturbatively in 2D). 
% \red{I understand your point, but I dont love the $\sim^\text{w}, \sim^\text{s}$ notations} \blue{[***I think that's fair. It's not a very informative notation. Do you have any alternatives?]}
% \blue{[x] Text cut}
%As mentioned before, the mechanism for this symmetry breaking is related to ground state selection in order-by-disorder phenomena \cite{Henley1989}, demonstrated rigorously in the RFO(2) model in 2D \cite{crawford2024random} and 3D \cite{crawford2013random,crawford2014random}).
More concretely, with probability one with respect to disorder realizations $\alpha$, 
% there is a region $D=D(\alpha)$ which has high density and such that for $r\in D$, 
the GSs $\phi(\br)$ is arbitrarily close to $\pm \pi/2$ on a highly dense subset of lattice sites, so that the disorder-averaged ordering $\overline{\bra e^{i\phi} \ket_{\alpha}}$ is arbitrarily close to $\pm i$ for sufficiently low temperatures \cite{crawford2013random,crawford2014random,crawford2024random}.
% , where $\overline{\cdots}$ denotes the average over disorder $\alpha$.  
% The physical picture here is that the spins are strongly ordered  ferromagnetically intra-layer, but the relative angle of ordering between the two layers is broken down into a $\dZ^2$ symmetry corresponding to the two possible orientations for the layer averages to point at right angles.

% The limit of strong disorder, $\ve \to \infty$, is more subtle.  
% The possible disorder-averaged GS can still be labeled by the angles $\sim^\text{s}\pm\pi/2,$ but  at first sight, the origin of this labeling appears  to be distinct from that in the weak disorder regime.\blue{More on this if space provides}
% To determine GSs, it is convenient to parametrize $\phi(\br) = \alpha(\br) +\delta \phi(\br).$ 

In the strong disorder limit $\ve \to \infty$, the GSs can still be labeled by $\phi \sim^\text{s}\pm\pi/2$, but the origin is distinct from that in the weak disorder.
To see this, it's instructive to assume the GS ansatz $\phi(\br) = \alpha(\br) +\delta \phi(\br)$.
Since the disorder strength is large compared to the in-plane interactions, $\ve \gg 1$, a GS must, first and foremost, attempt to minimize the inter-layer interaction in Eq. \eqref{eq:H}, leaving, presumably, small deviations $\delta \phi(\br) \ll 1$.  
Inserting this ansatz into Eq. \eqref{eq:H},
the effective inter-layer interaction is thus quadratic to lowest orders, i.e.,
$-\ve \cos(\phi(\br)-\alpha(\br)) \approx \ve \delta\phi(\br)^2/2$.
Writing $\theta^\pm =\theta\pm \phi/2$ where $\theta=(\theta^++\theta^-)/2$, the intra-layer interaction becomes
\begin{equation}
    \label{eq:H-rewrite}
    \sum_{\ell =\pm } \sH^{\text{XY}}(\theta^\ell) = -2\sum_{e} \cos\nabla \theta \cos \left(\frac{\nabla \phi}{2}\right)
    % =:\sH^{\text{XY}}(\theta, \phi)
\end{equation}
where the summation is over all edges $e =\br'\br$ and $\nabla \theta \equiv \theta(\br')- \theta(\br)$ is the lattice gradient.

%\red{Since $T\ll 1$ and the average phase $\theta$ is naively decoupled from the random phase $\alpha$,} 

Since $|\delta \phi| \ll 1 \ll h$, the in-plane interaction may be expanded up to linear order\footnote{The neglected terms shift GSs by higher orders in $1/h$.} in $\delta \phi$. 
Finally, if $T \ll 1$, we may reasonably assume that the \textit{average} phases $\theta$ are LRO (justification follows in the next section), which leads to the further simplification $\cos \nabla\theta \approx 1$.  
This leads to the \textit{effective} Hamiltonian, valid if $T\ll 1 \ll \ve,$ 
\begin{align}
    \label{eq:H-eff-zeroT}
    \sH_\alpha^\text{eff}(\delta \phi) 
    &= \sum_e \sin\left( \frac{\nabla \alpha}{2} \right)\nabla \delta \phi +\frac{\ve}{2} \sum_{\br} \delta \phi(\br)^2 \\
    % &= -\sum_{\br} \nabla_{\br}\cdot \sin\left( \frac{\nabla \alpha}{2} \right) \delta \phi(\br) \\
    % &\quad\:+\frac{\ve}{2} \sum_{\br} \delta \phi(\br)^2 \nonumber\\
    &= \frac{\ve}{2}\sum_{\br}\left[\delta \phi(\br) -\frac{1}{\ve} \nabla_{\br}\cdot \sin\left( \frac{\nabla \alpha}{2} \right) \right]^2
\end{align}
where equality follows from integration by parts and completing the square. 
Here $\nabla_{\br}\cdot$ denotes the lattice divergence operator at site $\br,$ and allows us to write the GS in the compact form
\begin{equation}
    \label{eq:GS}
    \delta \phi(\br) = \frac{1}{\ve} \nabla_{\br}\cdot \sin\frac{\nabla \alpha}{2}  \equiv \frac{1}{\ve}\sum_{\br' \sim \br} \sin\left( \frac{\nabla_{\br'\br} \alpha}{2} \right),
\end{equation}
where summation is over neighboring sites $\br'\sim \br$ in the graph $G$. 
% \blue{[***I reused the previous version since I thought it would be more concise to write the GS and how the divergence is defined in one line,  and because having an inline ratio $\frac{1}{h}$ never really sat quite right with me. It may also be nice to have the GS equation in display mode since it is used heavily in the figures and the next section and thus can be referenced.]}

% Here, we introduced the notation
% \begin{equation}
%     \label{eq:GS}
%      \frac{1}{\ve} \nabla_{\br}\cdot \sin\left( \frac{\nabla \alpha}{2} \right) \equiv \frac{1}{\ve}\sum_{\br' \sim \br} \sin\left( \frac{\nabla_{\br'\br} \alpha}{2} \right)
% \end{equation}
% with the summation over neighboring sites $\br'\sim \br$ in the graph $G$.  Thus the GS is straightforwardly given by $\delta \phi(\br) = \frac{1}{\ve} \nabla_{\br}\cdot \sin\left( \frac{\nabla \alpha}{2} \right).$ 

% \begin{equation}
%     \nabla_{\br}\cdot \sin\left( \frac{\nabla \alpha}{2} \right) = \sum_{\br' \sim \br} \sin\left( \frac{\nabla_{\br'\br} \alpha}{2} \right).
% \end{equation}
% The summation is over neighboring sites $\br'\sim \br$ in the graph $G$. 
% It's then straightforward that the GS is given by
% \begin{equation}
%     \label{eq:GS}
%     \delta \phi_\text{GS}(\br) = \frac{1}{\ve} \nabla_{\br}\cdot \sin\left( \frac{\nabla \alpha}{2} \right)
% \end{equation}

\begin{figure}[ht]
\subfloat[\label{fig:tildeZ2}]{%
  \centering
  \includegraphics[width=0.45\columnwidth]{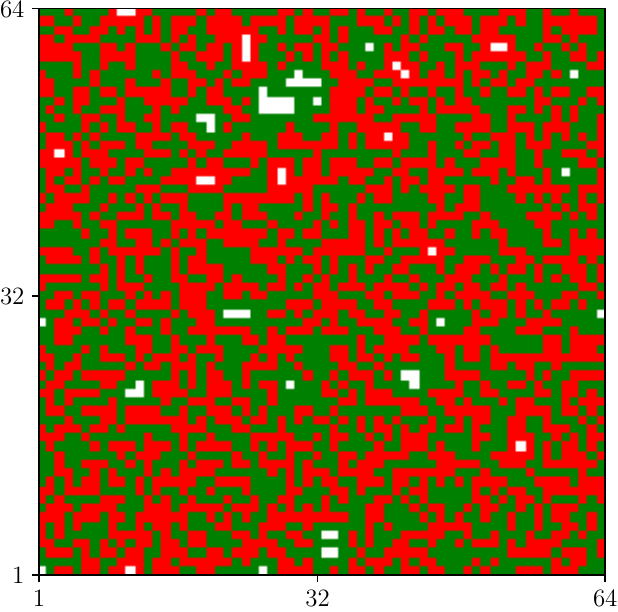}
}
\subfloat[\label{fig:Z2}]{%
  \centering
  \includegraphics[width=0.45\columnwidth]{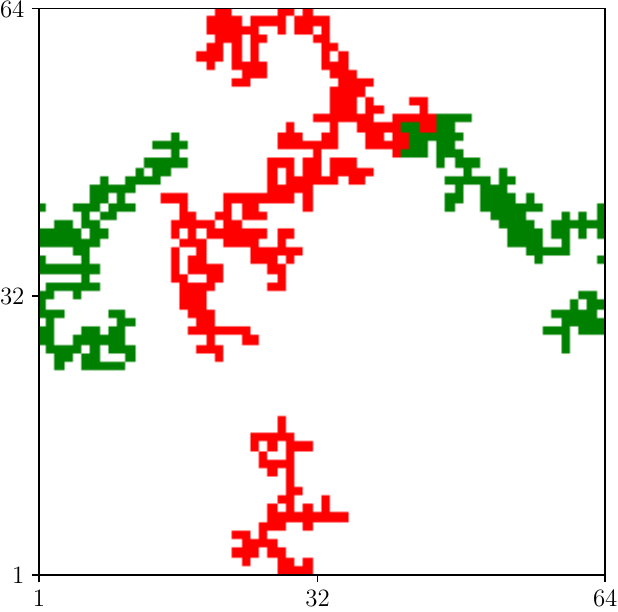}
}
\caption{Largest cluster.  For a given disorder realization $\alpha$ (with periodic boundary conditions), (a) plots the largest $\alpha=0$ cluster (green) and $\alpha =\pi$ cluster (red) on $\tilde{\dZ}^2$, while (b) plots those on $\dZ^2$. All other lattice sites are colored white. The difference between subplots (a), (b) is due to $p_c^\text{site}(\tilde{\dZ}^2) < 1/2 <p_c^\text{site} (\dZ^2)$
}
\label{fig:cluster}
\end{figure}

\textit{Emergence of a Fractal Structure} --- For a given sample of random phase shift $\alpha$, we partition the lattice sites into clusters\footnote{Neighboring sites with the same random phase $\alpha$ are connected.} of $\alpha=0,\pi$ (see Fig. \ref{fig:GS}).  
From Eq. \eqref{eq:GS}, if $\br$ is within the interior of a cluster, then $\delta \phi_\text{GS}(\br) = 0$ vanishes. 
However, if $\br$ is at a cluster boundary, then $\delta \phi_\text{GS}(\br) \gtrless 0$ with magnitude\footnote{Note that the GS solution is self-consistent with the ansatz assumption $\delta \phi \ll 1$} $\delta \phi_\text{GS} \sim 1/\ve$, $\ve\gg 1$.
In this sense, the ordering of the GS $\phi$ only occurs on $\alpha$-cluster boundaries, where $\phi$ uniformly slants in the $+\pi/2$ direction. (the alternative $\dZ_2$ degenerate GS $\phi \sim^\text{s} -\pi/2$ is such that $\phi$ slants in the $-\pi/2$ direction on the boundaries).

The key point is that, even though slanting is only present at boundaries between clusters, it is a bulk effect between a pair of \textit{infinite} components.  
Indeed, if $p_c^{\text{site}} <1/2,$ 
%at first glance, one might think that the average (bulk) magnetization scales like boundary/volume $\sim 1/L$ where $L$ is the linear system size, which vanishes in the limit $L\to \infty$.
% {\transparent{0.4} However, the $\alpha=0,\pi$ site percolations are not combinatorially dual and can therefore percolate simultaneously.} 
% \blue{[***I'm not sure if saying ``the site percolations are not combinatorially dual" provides more information than simplifying as follows]}
both the $\alpha=0,\pi$ sites percolate, each with a unique infinite cluster \cite{burton1989density}
and thus ergodicity dictates that the interface between the two infinite clusters is volume extensive (see Fig. \ref{fig:cluster}).

\textit{Comparison} --- It is worth reflecting on what distinguishes our \textit{bilayer} model from conventional \textit{single layer} models.
The \textit{single} layer RFO(2) model, with Hamiltonian
\begin{equation}
    \label{eq:H-single}
    \sH^\text{RFO(2)}_\alpha (\phi) = \sH^\text{XY}(\phi) -\ve\sum_{\br} \cos(\phi(\br)-\alpha(\br)),
\end{equation}
also has nontrivial GS $\phi\sim^\text{w} \pm \pi/2$ provided that $\ve$ is small \cite{dotsenko19812d,dotsenko1982spin,crawford2024random,crawford2013random,crawford2014random}.
Comparing with Eq. \eqref{eq:H-rewrite}, (assuming $\cos \nabla \theta =1$), the only difference is a factor of $1/2$ in the in-plane interactions.
Repeating the previous analysis in the regime $\ve \gg 1$ for $\sH^\text{RFO(2)}_\alpha$, the factor of $1/2$ would be absent from Eq. \eqref{eq:GS}, which implies a vanishing GS solution $\delta \phi =0$ since $\nabla \alpha =0, \pm \pi$.

\textit{Finite $T$ Effective Hamiltonian} --- Previously, we assumed that the average phase $\theta$ orders near zero temperature $T\ll 1$.
Let us justify this assumption by showing that $\theta$ undergoes a phase transition with finite transition temperature $\sim 1$.
% It's worth mentioning that the average phase $\theta \equiv (\theta^+ +\theta^-)/2$ is currently ill-defined as an order parameter since $\theta \mapsto \theta +\pi$ under $\theta^+ \mapsto \theta^+ +2\pi$ and $\theta^-\mapsto \theta^-$.
% One way of treating this is to use the following definition
% \begin{equation}
% \label{eq:PS}
%     \theta \equiv \frac{\theta^+ +\theta^-}{2} +\pi \mathbf 1\left\{\cos \left(\frac{\phi -\alpha}{2}\right) < 0 \right\},
% \end{equation}
% where $\mathbf 1\{A\}$ denotes the indicator function, i.e., $=1$ when condition $A$ is met and $=0$ otherwise.
% Note that if $\delta \phi = \phi-\alpha$ is within the $2\pi$-interval $(-\pi,\pi)$, then the definition reduces to the conventional definition. 
% In particular, since the system is in the strong disorder regime $\ve \gg 1$, the GS ansatz $\delta \phi \ll 1$ is still valid and thus there is no loss of generality when using the conventional definition of the average phase $\theta$. 
Repeating the previous expansion in Eq. \eqref{eq:H-eff-zeroT}, we find that the finite $T$ effective Hamiltonian is given by
\begin{align}
\label{eq:H-eff}
    \sH^\text{eff}_\alpha (\theta,\delta\phi) &= \sH^{\text{XY}}_\alpha(\theta) %-\sum_{e} \cos \nabla \theta \cos \left( \frac{\nabla\alpha}{2}\right) \\ 
    -\frac{1}{2\ve}\sum_{\br} \left[ \nabla_{\br} \cdot \sJ_\alpha(\theta) \right]^2 \nonumber\\
    &+\frac{\ve}{2} \sum_{\br}\left[\delta \phi(\br) -\frac{1}{\ve} \nabla_{\br} \cdot \sJ_\alpha(\theta) \right]^2,
\end{align}
where $(\sJ_\alpha)_{\br'\br}(\theta)=\cos(\nabla_{\br'\br} \theta)\sin(\nabla_{\br'\br} \alpha/2)$ and {
$\sH^\text{XY}_\alpha(\theta)$ originates from Eq. \eqref{eq:H-rewrite} with $\phi$ set to $\alpha$.}
The form of this Hamiltonian shows that  $\delta \phi$, conditional on $\theta$, is Gaussian distributed with mean $\ve^{-1} \nabla_{\br} \cdot \sJ_\alpha(\theta)$ and white noise covariance, and thus can be integrated out.   
The resulting  effective Hamiltonian, denoted $\sH^\text{eff}_\alpha(\theta),$ reduces to the first two terms in Eq. \eqref{eq:H-eff}.
In particular, the correlation functions in  $\phi$ can be computed using the effective Hamiltonian $\sH_\alpha^\text{eff}(\theta)$ alone, e.g., 
\begin{equation}
    \bra \sin \phi(\br)\ket_\alpha\approx \frac{1}{\ve}   \left\bra\nabla_{\br}\cdot\left[1\{\nabla \alpha \ne 0\}\cos\nabla\theta  \right] \right\ket^\text{eff}_\alpha,
    % \bra \sin \phi(\br)\ket\approx \frac{e^{i\alpha(\br)}}{\ve}   \left\bra\nabla_{\br}\cdot\left[\cos\nabla\theta  \sin\left( \frac{\nabla \alpha}{2} \right)\right] \right\ket^\text{eff},
\end{equation}
where $\bra\cdots\ket_\alpha,\bra \cdots \ket^\text{eff}_\alpha$ are with respect to $\sH_\alpha(\theta^\pm), \sH^\text{eff}_\alpha(\theta)$, respectively.

To zeroth order in $1/\ve$, $\sH^\text{eff}_\alpha(\theta) = \sH^\text{XY}_\alpha(\theta)$ describes a pair of independent XY models along distinct $\alpha=0,\pi$ clusters and thus $\theta$ can be naturally partitioned into $\vartheta^\pm$ by restricting to the distinct clusters.
% \begin{align}
%     \label{eq:H-eff-finiteT}
%     \sH^\text{eff}(\theta) &= -\sum_{e \sim_\alpha 0} \cos \nabla\theta -\sum_{e\sim_\alpha \pi} \cos\nabla\theta \\
%     &-\frac{1}{2\ve}\sum_{\br} \left[ \nabla_{\br} \cdot \left[\cos\nabla\theta  \sin\left( \frac{\nabla \alpha}{2} \right)\right] \right]^2, \nonumber
% \end{align}
% where $e\sim_\alpha 0$ denotes summing over edges $e=\br'\br$ within the same $\alpha =0$ cluster, i.e., $\alpha(\br')=\alpha(\br)=0$, and similarly define $e\sim_\alpha \pi$.
%To zeroth order in $1/\ve$, the effective Hamiltonian $\sH^\text{eff}(\theta)$ consists of two independent XY models defined on the $\alpha=0,\pi$ clusters and thus the spin configuration $\theta$ can be naturally partitioned into spins $\vartheta^\pm$, defined as the restrictions of $\theta$ on the $\alpha=0,\pi$ clusters.
If $p_c^\text{site} <1/2$, then both $\alpha=0, \pi$ subgraphs have unique infinite clusters \cite{burton1989density}. 
If $d\geq 3,$ the restrictions of $\vartheta^\pm$ to their respective infinite clusters  each undergo a phase transition at finite temperature (independent of $\ve$). 
Notably, this statement was recently demonstrated rigorously in Ref. \cite{dario2023phase}.

Thus,  for $T\ll 1$ (but independent of $h$), the $\alpha=0,\pi$ infinite clusters should be viewed as possessing internal LRO and each parameterized by a single $O(2)$ spin. 
It is then a question of how these effective degree of freedom orient with respect to one another.  
The latter is determined by the $O(1/h)$ term in $\sH_\alpha^\text{eff}(\theta).$ 
Note that it is  supported at the interface between distinct clusters and also depends only on the phase differences, i.e., $\varphi\sim \vartheta^+-\vartheta^-$.
Finally, the interaction is globally $\dZ^2$ invariant $\varphi \mapsto -\varphi$.
Hence, from symmetry considerations, $\sH^\text{eff}_\alpha(\theta)$ with strong disorder $\ve$ possesses the same features as a \textit{bilayer} Hamiltonian with a weak $1/\ve$ non-disordered $\cos^2 \varphi$ interaction, i.e.,
\begin{equation}
    \label{eq:H-clean}
    \sH^\text{clean}(\vartheta^\pm) = \sum_{\ell=\pm} \sH^\text{XY}(\vartheta^\ell) -\frac{1}{\ve} \sum_{\br} \cos^2 \varphi(\br)
\end{equation}
This correspondence is made rigorous in our detailed paper \cite{detailed}, and implies that in the limit $\ve \to \infty$, the disordered system in Eq. \eqref{eq:H} possesses a finite transition temperature if  the clean system in Eq. \eqref{eq:H-clean} possesses a finite transition temperature for arbitrarily weak interactions $1/\ve \to 0$.
The latter is then guaranteed by Ginibre's inequality \cite{ginibre1970general} on $\dZ^d$ for $d\ge 3$.

%\red{these two paragraphs need  work:}
% Moving on to the effect of the first order correction the term given $\br$, let $N_{\br}$ denote the set of neighbors of $\br$ so that $\alpha_{\br'}\neq \alpha_{\br}$.  Then the second line of \eqref{eq:H-eff-finiteT} may be rewritten as
% \begin{equation}
% \label{eq:1st}
%     -\frac{1}{2\ve}\sum_{\br}\left[\sum_{\br'\in N_{\br}} \cos\nabla_{\br\br'}\theta \right]^2.
% \end{equation}
% Using the formula $\cos(A)\cos(B)=\cos(A+B)+ \cos(A-B),$ it follows that this interaction falls within the class which satisfy the Ginibre inequalities.
% In particular, making use of the aforementioned work \cite{dario2023phase}, the pair of  infinite clusters corresponding to the $\alpha=\{0, \pi\}$ phase shift possess long range order internally.  But since \eqref{eq:1st} is volume-extensive between these two clusters, they are forced to either align or anti-align.

\textit{2D behavior} --- % In $\dZ^d, d\ge 3$, the Ginibre inequalities \cite{ginibre1970general} guaranty a finite ($\ve$-independent) transition temperature $T_\text{LRO}\sim 1$ for the clean Hamiltonian in Eq. \eqref{eq:H-clean}, below which the phase difference $\varphi$ possesses LRO.
%As discussed previously, this guarantees that the disordered bilayer model in Eq. \eqref{eq:H} is infinitely-stable in the strong disorder limit $\ve \to \infty$.
In $d=2$ with NN couplings, the previous argument fails because $p_c^\text{site}(\dZ^2) >1/2.$ This entails that the individual clusters are finite and exponentially decaying in their size distribution \cite{Grimmettbook}.
Adding NNN connections, so that $p_c^\text{site}(\tilde{\dZ}^2) < 1/2$, the above considerations only lead to quasi-LRO below a transition temperature  $T_\text{QLRO} \sim 1$.
Indeed, these conclusions are based on the fact that each infinite cluster on $\tilde{\dZ}^2$ exhibits a Kosterlitz-Thouless (KT) phase, rather than a true LRO phase. Whether  $\varphi$ has a LRO phase at $\ve$ independent temperatures remains an open question.
% To gain more insight into the behavior of the model on $\tilde{\dZ}^2,$ we further approximate  \eqref{eq:H-eff-finiteT} with the homogeneous bilayer Hamiltonian
% \begin{align}
%     \label{eq:H-clean}
%     \sH(\varphi) &= \sum_{\ell=\pm} \sH^\text{XY} (\varphi^\ell) - 1/\ve \sum_{\br} \cos (2\varphi).
% \end{align}
% where $\varphi=\varphi^+-\varphi^-.$
% The previous discussion entails that $\varphi$ possesses quasi-LRO below a finite $T_\text{QLRO} \sim 1$.
% Whether it possesses an $\ve$-independent LRO transition temperature $T_\text{LRO}$ is an open question.

\begin{figure}[ht]
\centering
\includegraphics[width=0.8\columnwidth]{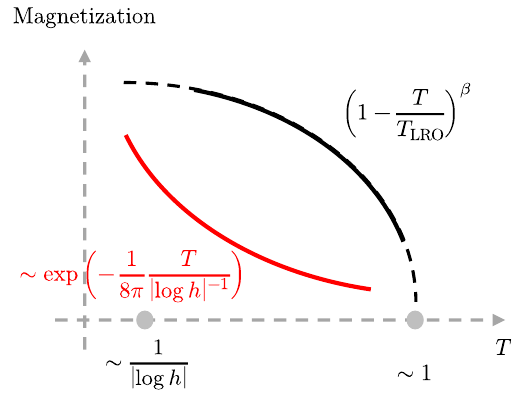}
\caption{Upper bound. The $x$ axis denotes the temperature, while the $y$ axis denotes an order parameter such as magnetization. The red line denotes the upper bound proved in Eq. \eqref{eq:upper-bound}. If $T_\text{LRO}\sim 1$, then universality implies that the magnetization follows the form of the black line with critical exponent $\beta$.
}
\label{fig:upper-bound}
\end{figure}

Indeed, field theory \cite{Seibergetal2021} suggests that the $\cos^2 \varphi$ interaction in Eq. \eqref{eq:H-clean} is always relevant, in the renormalization group sense,  and thus gives rise to LRO $T_\text{LRO} \sim 1$ in the limit $1/\ve \to 0$.
Numerics \cite{song2022phase,bojesen2014phase,maccari2022effects} corroborate this conjecture and show only a single phase transition for $\varphi$ -- a disordered phase at high $T$ and LRO below $T_\text{LRO} \sim 1$. 
% In particular, there is no observation of an intermediate massless phase between $T_\text{LRO}$ and $T_\text{QLRO}$ in contrast with the well known behavior for $\dZ_N$ clock models with $N\geq 5$.
Nevertheless, the mathematical evidence regarding the phase diagram is not clear cut.
For related $\dZ_2$ invariant systems in 2D, Ref. \cite{shlosman1980phase} shows that an analog to $\varphi$ exhibits LRO for temperatures below $\sim 1/\ve$ so that $T_\text{LRO}\gtrsim 1/\ve$ is bounded below\footnote{The proof utilizes reflection positivity and thus can be straightforwardly adapted to $\sH^\text{clean}$ in Eq. \eqref{eq:H-clean}}.  Our detailed paper \cite{detailed} proves that $\varphi$ can only order along $0,\pi$ (no ordering of $\sin \varphi$) and more interestingly,  for  $T\lesssim 1$
\begin{equation}
    \label{eq:upper-bound}
    |\bra \cos \varphi(\br) \ket| \lesssim \exp\left(-\frac{1}{8\pi} \frac{T}{|\log \ve|^{-1}}\right).
\end{equation}
% Notably, if  $1/\ve \ll 1$ so that there is a large gap between $T_\text{QLRO} \sim 1$ and $1/|\log\ve|$, 
Notably, if disorder $\ve$ is sufficiently large so that $1/|\log \ve| \ll 1 \sim T_\text{QLRO}$, 
then Eq. \eqref{eq:upper-bound} implies that if LRO occurs at $T_\text{LRO}\sim 1$, the magnetization must be exponentially small with respect to $|\log \ve|$.
In comparison (see Fig. \ref{fig:upper-bound}), universality conventionally implies that the magnetization $m$ 
% is of the form $(1-T/T_\text{LRO})^\beta$ for some critical exponent $\beta$ and thus 
becomes substantial at temperatures of the same scale as the transition temperature, i.e., $m \sim 1$ for $T\sim T_\text{LRO}$.

Based on the previous discrepancy, it appears that LRO occurs at $T_\text{LRO} \sim 1/|\log \ve|$, implying the possibility of an intermediate massless phase (for both the clean model in Eq. \eqref{eq:H-clean} and the disordered model in Eq. \eqref{eq:H}) in the limit $1/\ve \to 0$.
One may then speculate that such an intermediate phase has not been observed in numerics due to the extremely slow logarithmic decay.

\textit{Discussion and Outlook.} --- In this Letter, we restricted our attention to the scenario where the random phase $\alpha$ is equally likely to be $0,\pi$ (which corresponds to a vanishing disorder-average inter-layer interaction $\bar{J}=0$) and shown that the underlying reason for infinite (quasi-) stability is due to the random phase constituting a site percolation problem which satisfies $p_c^\text{site} < 1/2$.
The inequality suggests stability in parameter space, in the sense that if the random phase $\alpha$ is skewly-distributed so that $\alpha=0$ with probability $p$ (corresponding to a nonzero disorder-average $\bar{J}\ne 0$), then infinite (quasi-) stability still holds provided that $p_c^\text{site} < \min (p, 1-p)$.

\acknowledgements
We thank Steven A. Kivelson, Julian May-Mann, Akshat Pandey for helpful discussions.
ACY was supported by NSF grant No. DMR-2310312 at Stanford and the Laboratory for Physical Sciences at CMTC.  NC was supported by ISF grant No. 1557-21.

% \bibliography{refs.bib}
\bibliography{main.bbl}
\end{document}